\documentclass[seceq]{ptptex}

\usepackage{graphicx}

\title{
Dynamical vertex approximation
}

\subtitle{an introduction}    

\author{
K. \textsc{Held}$^1$, A. A. \textsc{Katanin} $^{2,3}$, and A. \textsc{Toschi}$^2$
}

\inst{
$^1$ Institute of Solid State Physics, Vienna University of Technology, 1040 Vienna, Austria\\
$^2$ Max-Planck-Institut f\"ur Festk\"orperforschung, 70569 Stuttgart, Germany
\\
$^3$ Institute of Metal Physics, 620219 Ekaterinburg, Russia
}

\recdate{Mmmmm DD, YYYY}

\abst{
We give an elementary introduction to a recent diagrammatic extension of
dynamical mean field theory (DMFT) coined dynamical vertex approximation (D$\Gamma$A). This approach contains the important local correlations of DMFT, giving, among others, rise to quasiparticle renormalizations, Mott-Hubbard transitions and magnetism, but also non-local correlations beyond. The latter  are  at the very essence
of many physical phenomena in  strongly correlated elecectron systems.
As correlations are treated equally on all length scales,
D$\Gamma$A allows us to describe physical phenomena such as magnons, quantum criticality, and the interplay between antiferromagnetism and superconductivity.
We review results hitherto obtained for the Hubbard model in dimensions $d=3$, 2, and 1.
}

\begin{document}

\maketitle

\section{Introduction}
Dynamical mean field theory (DMFT) \cite{DMFT,DMFT2,DMFTREV,DMFTPT}
has been a success story
for the theoretical modeling of
strongly correlated electrons systems since it contains the major contribution
of electronic correlations: the local ones.
Diagrammatically, DMFT  corresponds  to
all topologically distinct Feynman diagrams, albeit only
for their local part. Hence, it is a non-perturbative approach
which, from another point of view, can be envisaged  as a mean-field theory where
the local Coulomb interaction $U$
on all sites but one is replaced by a dynamical (frequency-dependent) mean field
given in terms of the self energy $\Sigma$.
First, DMFT was applied to model Hamiltonians for  studying, among others, 
 the  Mott-Hubbard transition \cite{DMFT2,DMFTREV,MHT},
magnetism \cite{MAG1,MAG2,MAG3,MAG4,MAG5},
and the strong quasiparticle renormalization of  correlated electrons, including
the only recently observed kink in the quasiparticle's dispersion relation \cite{kink1,kink2,kink3}.
In the last years,
the focus shifted
towards the realistic calculation of material properties
\cite{LDADMFT1,LDADMFT2,LDADMFT3,LDADMFT4,LDADMFT5} and the inclusion of non-local correlations,
which are the topic of our paper.

While usually smaller in absolute terms than the local correlations,
the  non-local correlations nonetheless become essential
at low temperatures where they lead to pertinent correlation effects
such as spin fluctuations, magnons, quantum criticality, 
and (possibly) superconductivity. Actually, many of the most 
fascinating phenomena of strongly correlated electron systems 
are driven by non-local correlations.

\begin{figure}[tb]
\begin{center} 
\includegraphics[clip=true,width=0.7\textwidth]{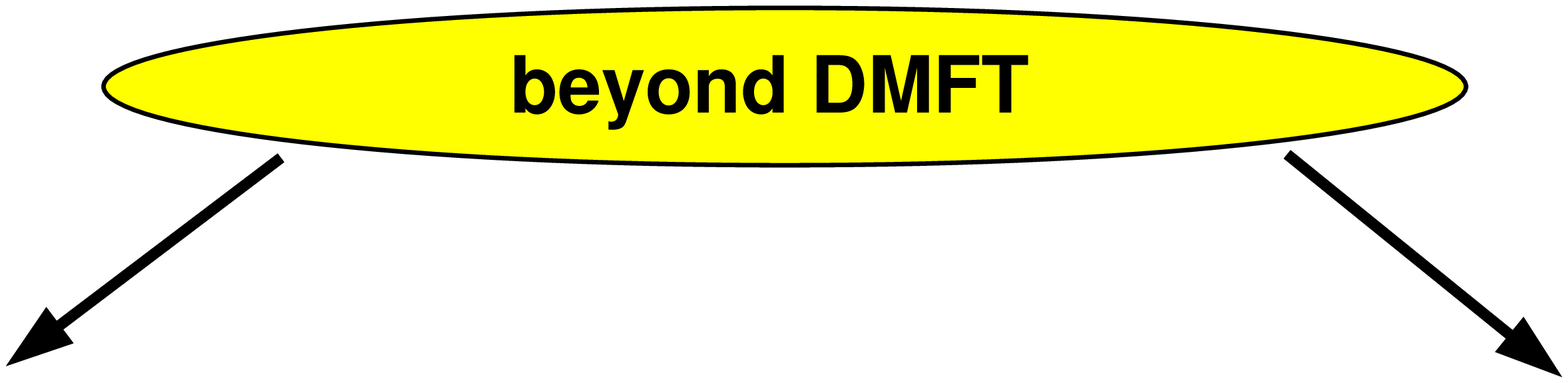}\hspace{1.cm}\phantom{a}
\end{center}
{\bf \large cluster extensions of DMFT}
\hfill
{\bf \large diagrammatic extensions of DMFT}
\vspace{.2cm}

\includegraphics[clip=true,width=0.4\textwidth]{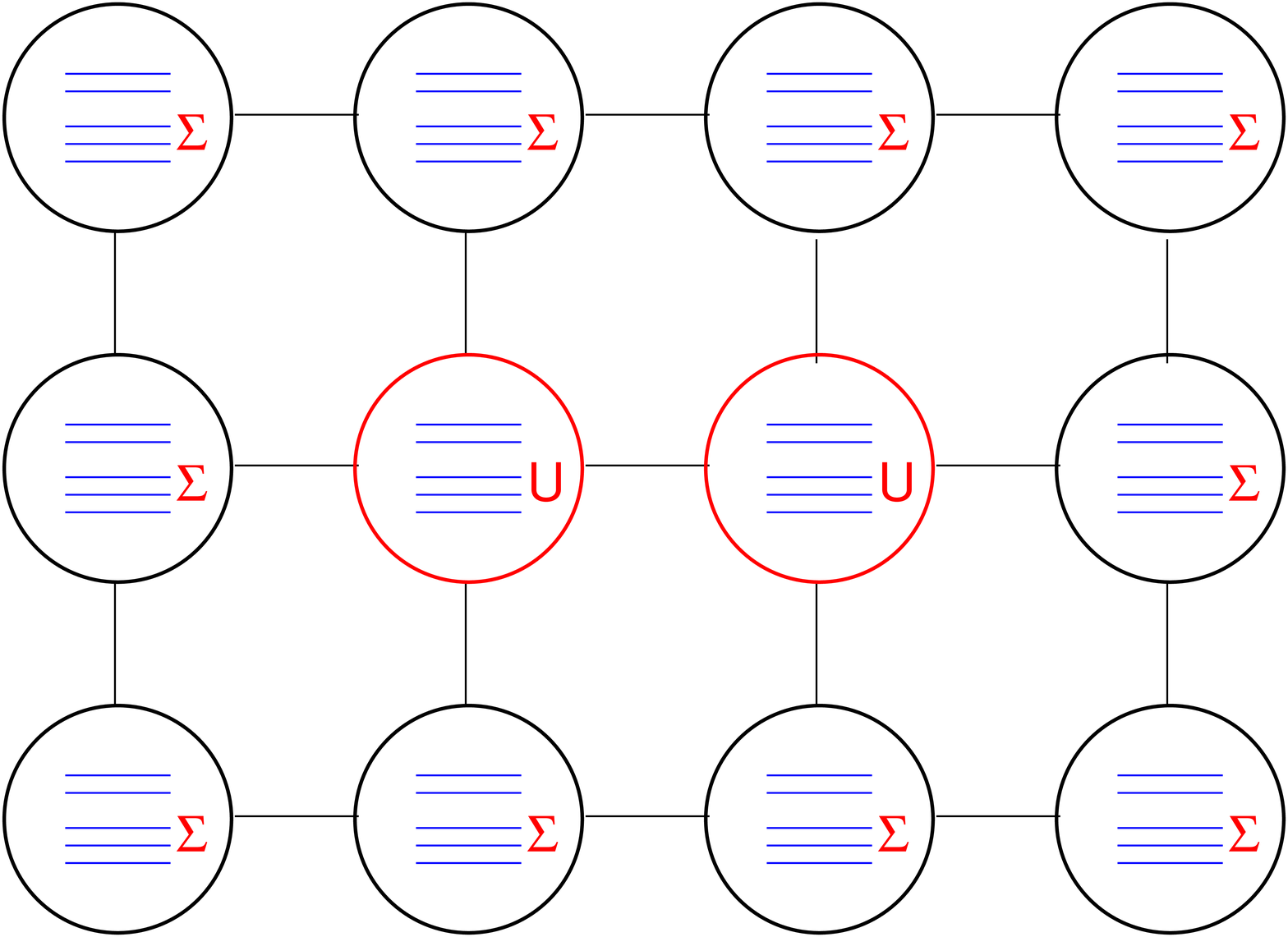}
\hfill 
\includegraphics[clip=false,width=0.52\textwidth]{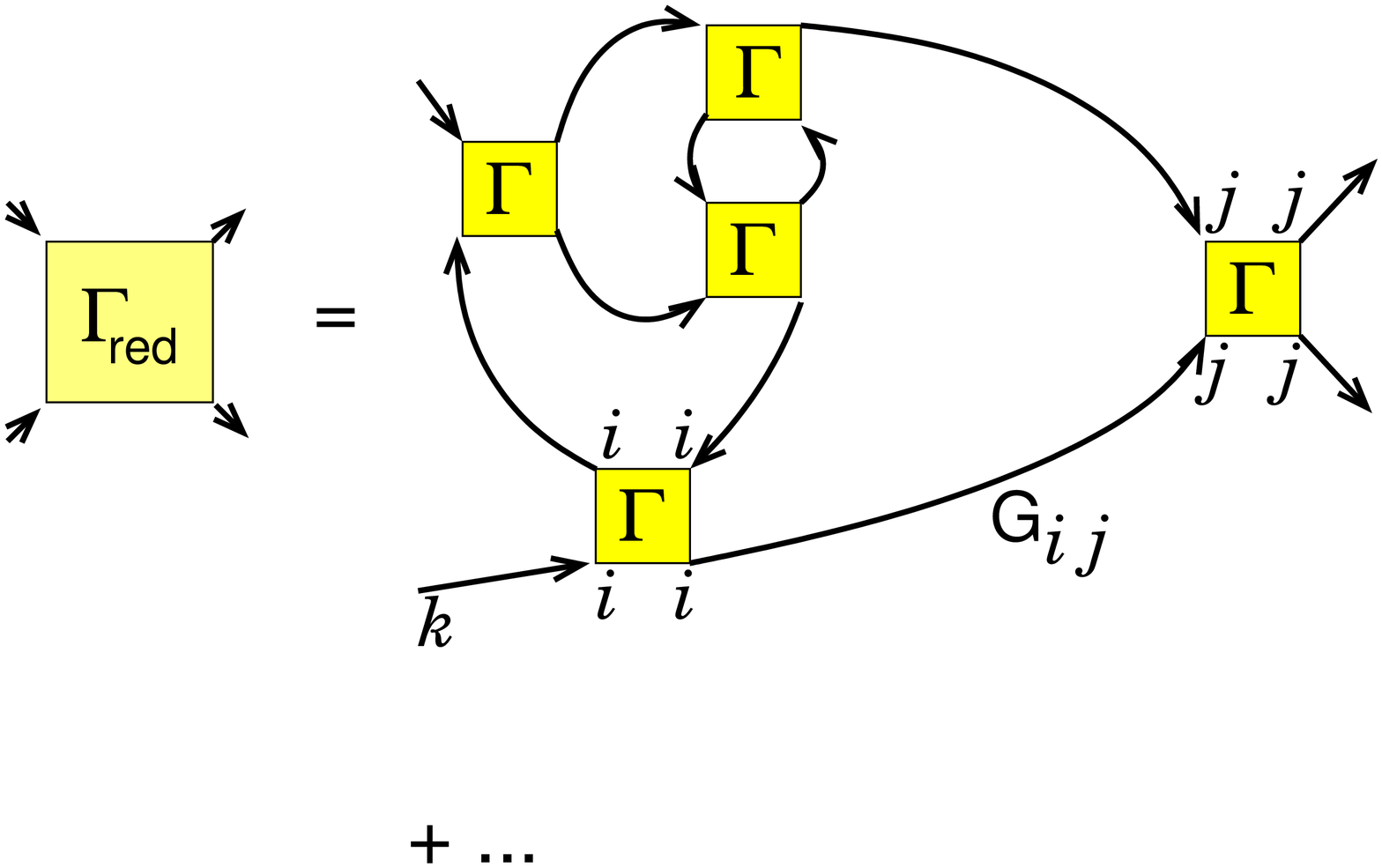}

\caption{Two ways to include non-local correlations beyond DMFT.
In cluster extensions of DMFT (left), a couple of sites  (here two)
are embedded in a dynamical mean field which can be formulated through
the  self energy $\Sigma$. Restricted to a single site, DMFT
is recovered.
In diagrammatic extensions, the local contribution of DMFT
is supplemented by certain non-local contributions.
Shown is the dynamical vertex approximation (D$\Gamma$A) 
where the fully irreducible vertex is purely local.
These local building blocks are connected by non-local
Green function lines $G_{ij}$, resulting in a non-local
reducible vertex and self energy. Reducing $G_{ij}$ to its local contribution
$G_{ii}$ reproduces DMFT.
\label{Fig:2ways}}
\end{figure}

Two different paths have been followed to extend DMFT by non-local
correlations, see Fig.\ \ref{Fig:2ways}.
Cluster extensions consider instead of the single DMFT site 
 (the local problem or Anderson impurity model) a couple of sites and take correlations 
between these sites into account, see Ref.\ \citen{review} for a review.
There are two main variants: cellular (or cluster) DMFT \cite{clusterDMFT,clusterLK,clusterDMFT2} where the self
energy is directly related to a supercell in real space, and the
dynamical cluster approximation (DCA) \cite{DCA1,DCA2,DCA3,clusterLK}
where it is patched in ${\mathbf k}$ space.
A generalization in form of a self energy functional approach 
has been formulated more recently by Potthoff \cite{Potthoff03a,Potthoff06a}.
Also
the $1/d$
expansion of DMFT \cite{Schiller,Zarand00} can be 
subsumed here since it connects
a single-site cluster and a two-site cluster.
These approaches have been applied successfully for studying
the one-dimensional \cite{Capone04a,clusterDMFT3} and two-dimensional  \cite{DCASC,clusterLK,Maier05a,Arita06a,review}
Hubbard model, as well as
for the investigation of spin-Peierls physics 
in Ti$_2$O$_3$\cite{Poteryaev04a}, NaV$_2$O$_5$ \cite{Mazurenko02a}, and  VO$_2$ \cite{Biermann05a}. The biggest drawback of the cluster extensions
is their restriction to short-range correlations because the
numerical effort is exponentially growing with the number
of cluster sites, similar to the direct quantum Monte Carlo simulation\cite{BSS}
of the lattice problem. For the one-band Hubbard model, approximately 30-60 sites
are possible which means a correlation length of about 3 sites in two dimensions and 1 site in three dimensions.
In realistic multi-orbital calculations, the effort increases further with the number of orbitals.
Hence, hitherto only two sites 
have been considered.

\begin{figure}[tb]
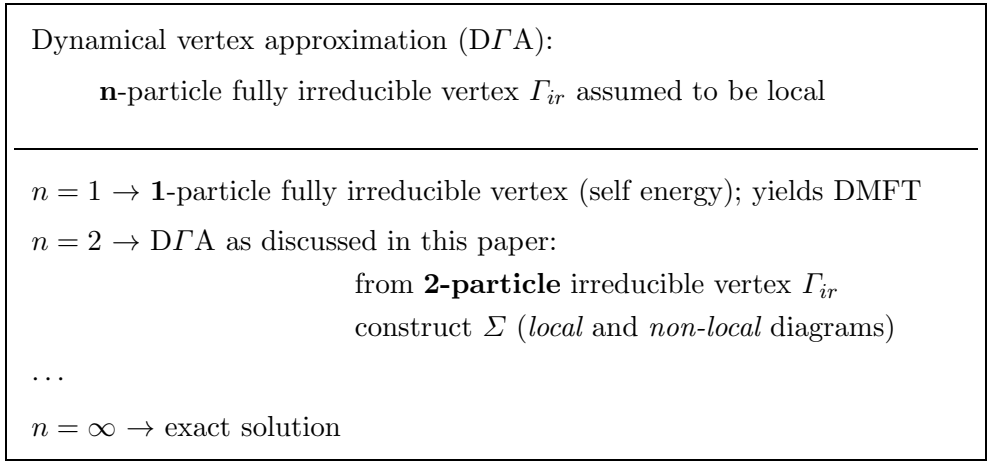


\vspace{.2cm}

 \begin{center}
\fbox{{\parbox{12.8cm}{

\vspace{.2cm}

\hspace{.1cm} Dynamical vertex approximation (D$\Gamma$A):

\vspace{.2cm}

\hspace{1cm}  $\mathbf n$-particle fully irreducible vertex  $\Gamma_{ir}$ assumed to be local
\vspace{.2cm}

\hrulefill
\vspace{.2cm}

\hspace{.1cm} $n=1$ $\rightarrow$  $\mathbf 1$-particle fully irreducible vertex (self energy); yields DMFT
\vspace{.2cm}

\hspace{.1cm} $n=2$ $\rightarrow$ {D${\Gamma}$A} as discussed  in this paper:

\vspace{.1cm}

\hspace{4.399cm} from {\bf 2-particle} irreducible vertex {$\Gamma_{ir}$}

\vspace{.1cm}

\hspace{4.399cm} construct {$\Sigma$}  ({\em local} and {\em non-local} diagrams)

\vspace{.2cm}

\hspace{.1cm} $\cdots$

\vspace{.2cm}

\hspace{.1cm} $n=\infty$ $\rightarrow$ exact solution

\vspace{.2cm}
}}}
\vspace{.2cm}

\end{center}
\caption{D$\Gamma$A as the natural extension of DMFT to the $n$-particle fully irreducible vertex.
\label{Fig:DGAgeneral}}
\end{figure}

Obviously, we need another route to tackle \emph{long-range} correlations. This is possible through diagrammatic  extensions
of DMFT which were addressed recently by serveral groups. Kuchinskii \textsl{et al.} \cite{Sadovskii05,Sadovskii05b}
combined the local DMFT self energy of the Hubbard model
with the non-local self energy of the
spin-Fermion model and included long-range correlations this way.
 Rubtsov \emph{et al.}
\cite{DualFermion} proposed a dual Fermion approach to this end.
Tokar and Monnier \cite{Tokar07a} developed a perturbative extension which however only includes
short-range correlations.
We think  that a natural extension of DMFT coined 
 dynamical vertex approximation (D$\Gamma$A) \cite{DGA}, which also
includes the spin-Fermion diagrams in a systematic way \cite{DGA3},
is particularly promising. Instead of assuming the one-particle fully irreducible
vertex,
i.e., the self energy, to be purely local as in DMFT, D$\Gamma$A 
assumes the same for the $n$-particle fully irreducible vertex $\Gamma_{ir}$ \cite{DGA}, see Fig.\ \ref{Fig:DGAgeneral}.
These local building blocks  are then connected by non-local Green functions, yielding a non-local reducible vertex and self-energy; see diagrams in the right panel of Fig.\ \ref{Fig:2ways}. Most natural (and feasible) is the case $n=2$, i.e., the two-particle
vertex. But, in principle, implementations with higher $n$ are possible
and the exact solution is recovered in the limit $n\rightarrow \infty$.
The D$\Gamma$A has been proposed by Toschi {\em et al.}
\cite{DGA} and independently (with minor differences) by Slezak  {\em et al.} \cite{DGA2},
who employ a DCA cluster as a starting point and
a perturbative calculation of the one-frequency vertex.
Very similar ideas including less diagrams have also been 
put forward independently by 
Kusonose \cite{Kusunose}.

The paper is organized as follows:
Section \ref{Sec:DGA} focuses on the D$\Gamma$A approach and algorithm, introducing
the full version of the approach in Section \ref{Sec:fullDGA},
a restriction to the most important diagrams for magnetic fluctuations 
in Section \ref{Sec:phDGA}, and  a Moriyaesque $\lambda$ correction
in Section \ref{Sec:DGAlambda}. Physical results
for the Hubbard model
in $d=3$, 2, and 1
are presented
in Sections  \ref{Sec:3d},  \ref{Sec:2d},  and \ref{Sec:1d},
respectively. Finally, 
 Section \ref{Sec:summary} gives a summary and outlook.

\section{D$\Gamma$A algorithm}
\label{Sec:DGA}
\subsection{Full (self-consistent) D$\Gamma$A}
\label{Sec:fullDGA}

\begin{figure}[tb]
\begin{center}
\includegraphics[width=7.1cm]{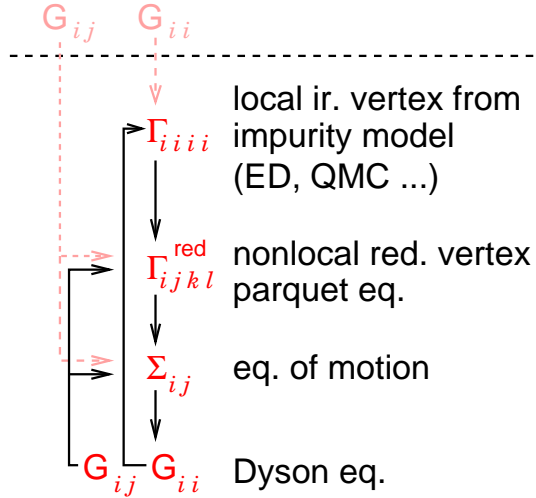}
\end{center}
\caption{Full (self-consistent) D$\Gamma$A algorithm.
\label{Fig:DGA}}
\end{figure}

Taking the locality of the fully irreducible\footnote{That is, cutting two Green function lines of the Feynman diagram for the vertex does not divide the vertex into two parts.} vertex as a starting point,
we can formulate
the D$\Gamma$A approach as following, see  Fig.\ \ref{Fig:DGA} for the algorithm:
Starting with an arbitrary local Green function,
we first have to calculate the fully irreducible vertex.
This is well defined diagrammatically, but in practice one will solve
the Anderson impurity model numerically using, for example,
the exact diagonalization (ED) \cite{DGA} or quantum Monte Carlo simulation (QMC)
in its Hirsch-Fye
\cite{
Hirsch86a}, projective \cite{Feldbacher06}, continuous time 
\cite{Rubtsov04a, Werner06a} or hybrid \cite{Sakai06a} variant.
In these numerical approaches, one actually does not compute  the irreducible vertex  directly, but instead the local three-frequency (generalized) spin and charge 
susceptibility and from this --through the local 
parquet equations \cite{Dzy,parquet,Janis2}--
the fully irreducible vertex; see Ref.\ \citen{DGA} for details.

In a second step, the (non-local) reducible vertex is calculated 
from  this (local) fully irreducible vertex. Fig.\ \ref{Fig:2ways}
shows as an example one reducible  diagram constructed from the
irreducible vertex. But of course, we have to sum over  all reducible 
Feynman diagrams  \footnote{That is,  all Feynman diagrams, which have the local irreducble vertex and the 
non-local Green function as building blocks, and which do not separate into two parts when cutting two Green function lines.}. This can be achieved more elegantly through the parquet
equations \cite{Dzy,parquet,Janis2}, which is still 
a challenge for an actual implementation of this full D$\Gamma$A
approach.

\begin{figure}[tb]
\begin{center}
\includegraphics[width=6.7cm]{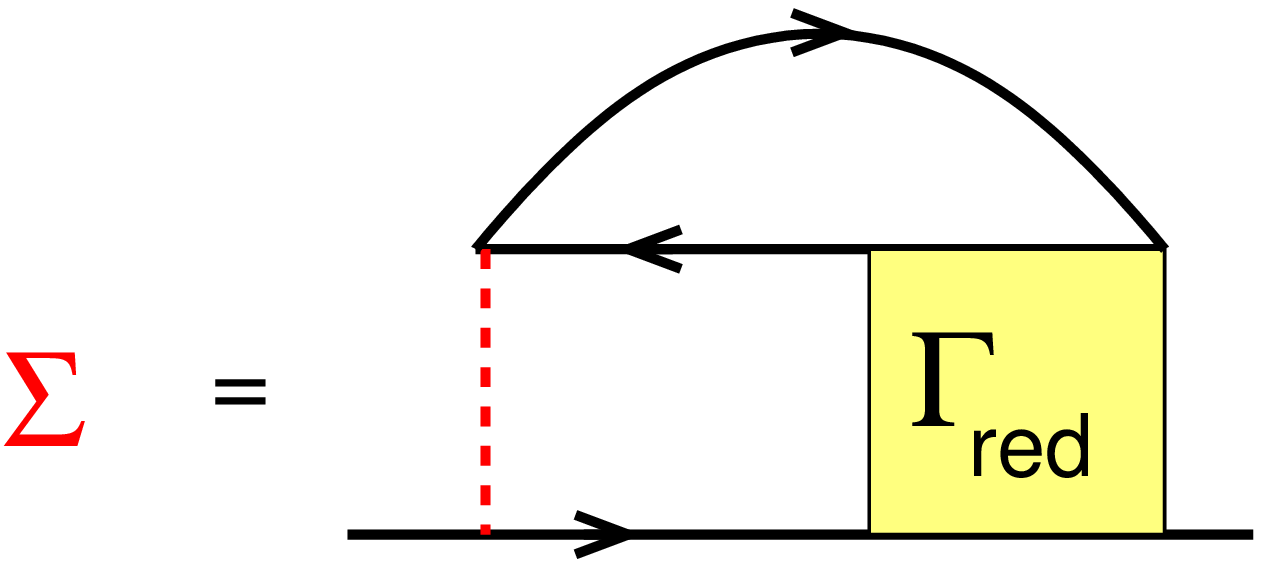}
\end{center}
\caption{Diagrammatic representation
of the equation of motion connecting reducible vertex and 
self-energy. \label{Fig:DGAsigma}}
\end{figure}

Having the non-local reducible vertex, one can easily obtain the 
non-local D$\Gamma$A self energy (third step in Fig.\ \ref{Fig:DGA})
 through an exact equation which follows from the Heisenberg equation of motion \cite{parquet,Janis2}:
\begin{eqnarray}
\Sigma _{\mathbf{k},\nu} \!=\!U \frac{n}{2}\! -T^2U\!\sum_{\stackrel{\scriptstyle \nu{^{\prime }}\omega}{\mathbf{k^{\prime}\mathbf{q}}}}\Gamma _{\mathbf{k
k{^{\prime}}q}}^{\nu \nu ^{\prime } \omega\uparrow\downarrow }G_{\mathbf{k}^{\prime }\mathbf{\!+q},\nu
^{\prime }\!+\omega }G_{\mathbf{k}^{\prime }\!,\nu ^{^{\prime }}}G_{\mathbf{k\!+q}%
,\nu +\omega };
\label{Eq:DGA}
\end{eqnarray} 
see Fig.\ \ref{Fig:DGAsigma}
for the diagrammatic representation.
Here, $\mathbf k$, $k^{\prime}$, and $\mathbf{q}$ denote the three involved
wave vectors;
$\nu$, $\nu ^{\prime }$, and $\omega$ the corresponding   Matsubara frequency; $n$ the number of electrons per site, $U$ the interaction and $T$ the temperature;
$\Gamma _{\mathbf{k k{^{\prime}}q}}^{\nu \nu ^{\prime }}$ is the afore calculated reducible 
four-point vertex and $G_{\bf{k},\nu}$ the Fourier-transform
of the  non-local Green function $G_{ij}$ taken as a starting point
in the first line of Figures \ref{Fig:DGA}.

From the self energy Eq.\ (\ref{Eq:DGA}),
the  dispersion relation $\epsilon_{\bf{k}}$ of the lattice and the
chemical potential $\mu$,
we can now calculate a new local and non-local
 Green function via the Dyson equation
\begin{equation}
G_{\bf{k},\nu}=[i \nu_n-\epsilon_{\bf{k}}+\mu-\Sigma_{\bf{k},\nu}]^{-1}
\label{Eq:GF}.
\end{equation}
After this fourth step of the algorithm,
 we can go back to step 1 and iterate the D$\Gamma$A 
algorithm Fig.\ \ref{Fig:DGA} until  convergence is reached.

\begin{figure}[tb]
\begin{center}
\includegraphics[width=2.9cm,angle=270]{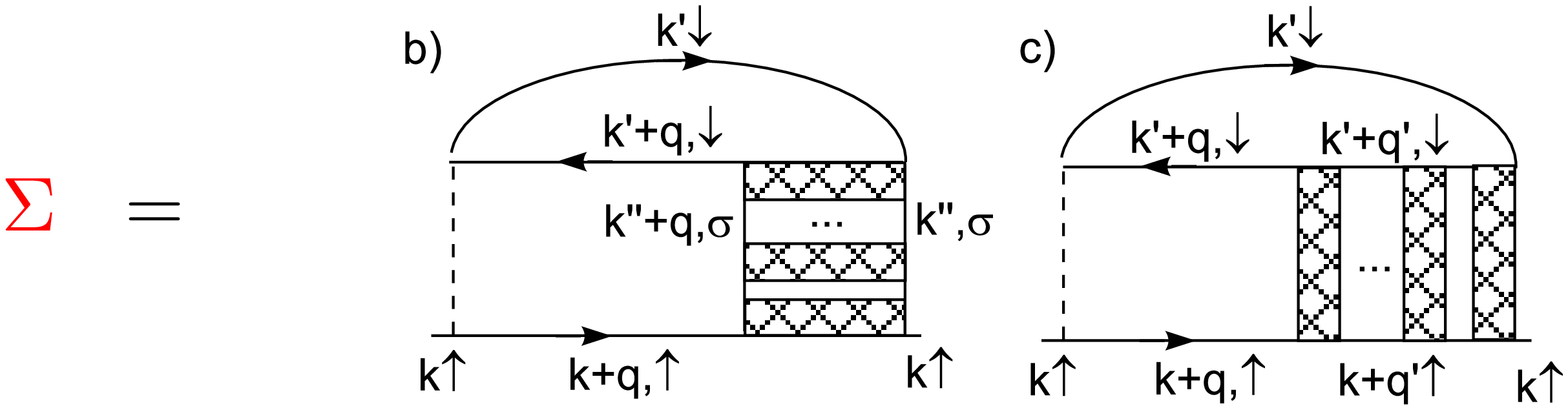}
\end{center}
\caption{
 The two particle-hole
channels contributing to the self-energy
in the ladder approximation, in the longitudinal and transversal spin channel, respectively. Instead of the
bare interaction, ladder diagrams are constructed from
the local vertices (crosshatched) irreducible in the corresponding spin and charge channels.
[reproduced from Ref.\ \citen{DGA}]}
 \vspace{-3mm}
\label{fig1}
\end{figure}

\subsection{Restriction  to the two-particle-hole channels}
\label{Sec:phDGA}
Instead of this full (self-consistent) D$\Gamma$A approach, we have, in a first step,
implemented a simplified scheme \cite{DGA}: Neglecting the particle-particle 
channel of the Parquet diagrams, we
restricted ourselves 
to the two particle-hole channels which are longitudinal and transversal, respectively,
 in the spin indices; see Fig. \ref{fig1} b) and c).
This  restriction is reasonable for spin fluctuations where these channels are dominating
and actually diverging close to the antiferromagnetic phase transition
in two dimensions. For other physical phenomena such as  superconductivity
the particle-particle channel is, as a matter of course, essential.

\begin{figure}[tb]
\begin{center}
\includegraphics[width=7.1cm]{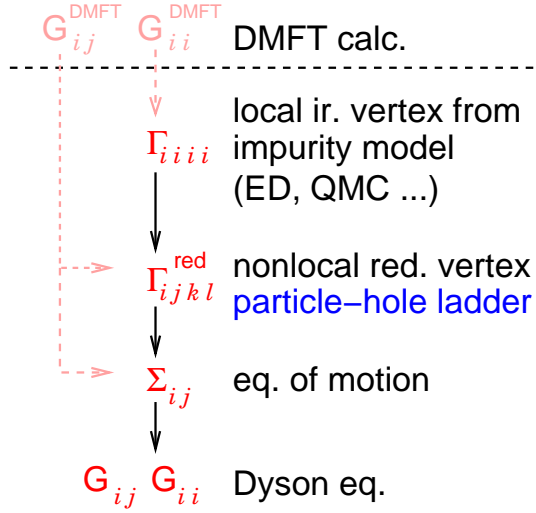}
\end{center}
\caption{Restriction of the D$\Gamma$A algorithm to particle-hole ladder diagrams.
\label{Fig:DGA2}}
\end{figure}

The algorithm is shown in  Fig.\ \ref{Fig:DGA2}:  We first calculate the irreducible vertex $\Gamma _{s (c),\text{ir}}$ in the
spin and charge channel
from the numerically obtained susceptibility
of the Anderson impurity model (step one in Fig.\ \ref{Fig:DGA2}), see Eqs. (9)-(11) of Ref.\ \citen{DGA}.
From this we calculate the reducible vertex $\Gamma _{s(c),\mathbf{q}}^{\nu\nu^{\prime } \omega }$ (step two  in Fig.\ \ref{Fig:DGA2})
through the ladder summation which yields
\begin{eqnarray}
\Gamma _{s(c),\mathbf{q}}^{\nu\nu^{\prime } \omega }=[(\Gamma _{s (c),\text{ir}}^{\nu
\nu^{\prime } \omega })^{-1}-\chi _{0\mathbf{q} \omega }^{\nu ^{\prime }}\delta _{\nu \nu
^{^{\prime }}}]^{-1}
\label{Eq:Gamma}
\end{eqnarray}
where
$\chi _{0\mathbf{q}\omega}^{\nu ^{\prime }}=-T\sum_{\mathbf{k}}G_{\mathbf{k},\nu ^{\prime } }G_{\mathbf{k}+%
\mathbf{q},\nu ^{\prime } +\omega }$ is the bubble contribution to the susceptibility.
In the third step of  Fig.\ \ref{Fig:DGA2},
the self energy is obtained from these vertices through
Eq.\ (\ref{Eq:DGA}) which  yields
\begin{eqnarray}
\Sigma _{\mathbf{k},\nu } &=&
\frac{1}{2}{Un}+\frac{1}{2}TU\sum\limits_{\nu ^{\prime }\omega ,\mathbf{q}%
}\chi _{0\mathbf{q}\omega }^{\nu ^{\prime }}\left( 3\Gamma _{s,%
\mathbf{q}}^{\nu\nu ^{\prime } \omega }-\Gamma _{c,\mathbf{q}}^{\nu
\nu ^{\prime } \omega }\right.   \nonumber \\
&&\left. +\Gamma _{c,\text{loc}}^{\nu\nu ^{\prime } \omega }-
\Gamma _{s,\text{loc}}^{\nu\nu ^{\prime } \omega }\right) G_{\mathbf{k+q}%
,\nu +\omega }.
\label{Eq:final}
\end{eqnarray}
for the restriction to the ladder diagrams in the two particle-hole channels
of Fig.\ \ref{fig1} b) and c), respectively. Here, $\Gamma _{s(c),\text{loc}}$ is the local ($\mathbf q$-summed) counterpart of $\Gamma _{s(c),\mathbf{q}}$.
With the restriction to the two particle-hole channels, actually  a non-self consistent calculation
is preferable so that the algorithm ends with a calculation of the D$\Gamma$A
Green function from the self energy of Eq.\ (\ref{Eq:final}), using the Dyson Eq.\ (\ref{Eq:GF}).

\subsection{Moriyaesque $\lambda$ correction}
\label{Sec:DGAlambda}
\begin{figure}[tb]
\begin{center}
\includegraphics[width=7.1cm]{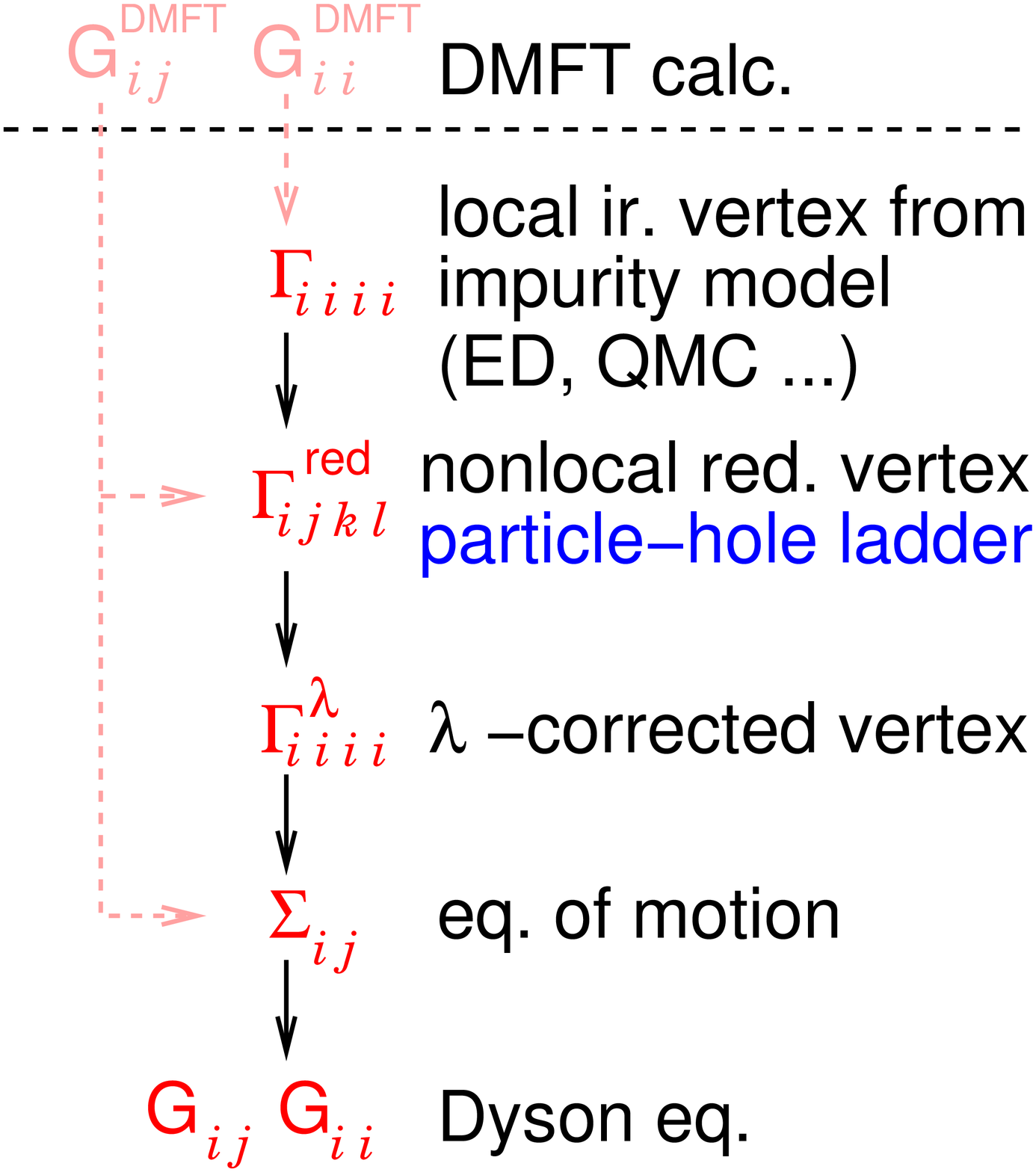}
\end{center}
\caption{D$\Gamma$A algorithm with Moriyaesque $\lambda$ correction.
\label{Fig:DGA3}}
\end{figure}

As a numerically inexpensive alternative to the fully self-consistent D$\Gamma$A algorithm,
we recently introduced a Moriyaeque $\lambda$ correction \cite{DGA3} to the 
algorithm of Section \ref{Sec:phDGA}. In the Moriya theory of weak
itinerant magnets \cite{Moriya}, the perturbatively obtained
susceptibility is corrected as follows
\begin{equation}
\chi _{\mathbf{q}\omega }^{s}\longrightarrow \left[ (\chi _{\mathbf{q}\omega
}^{s})^{-1}+\lambda\right] ^{-1},  \label{his}
\end{equation}%
where $\chi _{\mathbf{q}\omega }^{s}=\sum_{\nu \nu^\prime} \chi^{\nu \nu^\prime}_{s \mathbf{q}}$ is the susceptibility.
We do the same
but for strongly correlated electron systems
in the $\lambda$-corrected D$\Gamma$A. In our approach, the value of $\lambda$ is fixed by the sumrule%
\begin{equation}
-\int_{-\infty }^{\infty }\frac{d\nu }{\pi }\mbox{Im}\Sigma _{\mathbf{k},\nu
}=U^{2}n(1-n/2)/2.
\end{equation}
Eq.\ (\ref{his}) translates to a $\lambda$ correction
of the vertex via
\begin{eqnarray}
\chi _{s(c),\mathbf{q}}^{\nu \nu ^{\prime }\omega }&=&[(\chi _{0\mathbf{q}%
\omega }^{\nu ^{\prime }})^{-1}\delta _{\nu \nu ^{^{\prime }}}-\Gamma _{s(c),%
\text{ir}}^{\nu \nu ^{\prime }\omega }]^{-1}
\end{eqnarray}%
see Ref.\ \citen{DGA2} for details.
Hence in the  D$\Gamma$A algorithm, the $\lambda$ correction is one
additional step, the third step of  Fig.\ \ref{Fig:DGA3}.
Otherwise the algorithm remains unchanged.

Through the $\lambda$ correction, we include  important  effects
of the self-consistent full D$\Gamma$A. In particular, the reduction of the magnetic
transition temperature. In two dimensions, even the Mermin-Wagner theorem is fulfilled
\cite{DGA3}.

\section{Results for the Hubbard model}
\subsection{Three-dimensional Hubbard model}
\label{Sec:3d}

\begin{figure}[t]
\begin{center}
\includegraphics[width=12.cm]{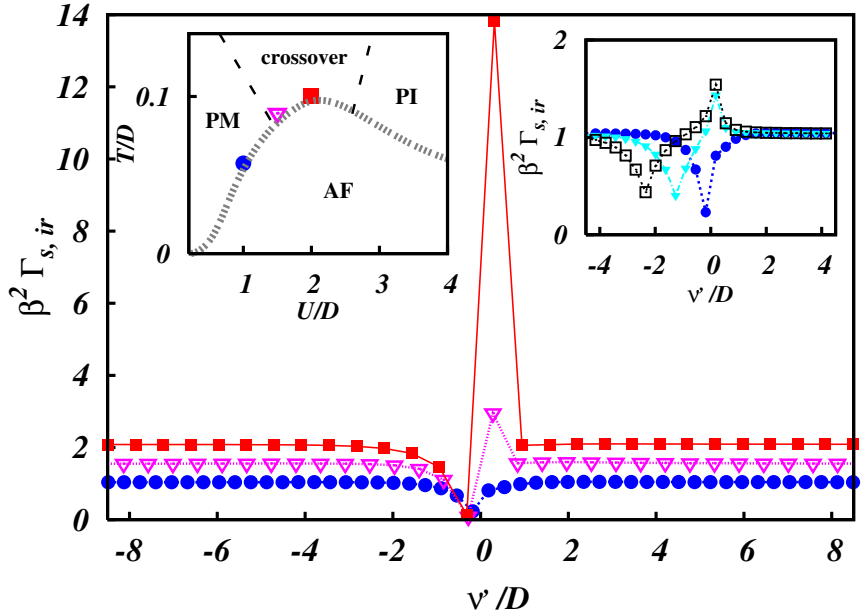}
\end{center}
\caption{Dependence of the local irreducible vertex
${\Gamma_{s,ir}^{\nu^{\prime }\nu=\pi\!/\!\beta\;
\omega=0 }}$ on the incoming Fermionic frequency 
${\nu^{\prime }}$,
for the three different values of  $U$ and $T$ indicated
as symbols in the left inset, which shows
the DMFT phase diagram with paramagnetic metallic (PM), 
insulating (PI), and antiferromagnetic (AF) phase as well as the crossover region between PM and PI.
Right inset: Same as main panel 
but at fixed $U=1D$ and  at different $\omega$'s  (squares: $\omega=12 \pi T$;
 triangles: $\omega=6 \pi T$; circles $\omega=0$). [reproduced from Ref.\ \citen{DGA}]}
\label{fig3}
\label{Fig:3d}
\label{Fig:3D}
\end{figure}

Let us turn to the results, starting with the three-dimensional 
Hubbard model on a cubic lattice
\begin{equation}
H=-t\sum_{\langle ij\rangle \sigma }c_{i\sigma }^{\dagger }c_{j\sigma
}+U\sum_{i}n_{i\uparrow }n_{i\downarrow }.  \label{H}
\end{equation}%
Here, $t$ denotes the hopping amplitude between nearest-neighbors, $U$ the
Coulomb interaction, $c_{i\sigma }^{\dagger }$($c_{i\sigma }$) creates
(annihilates) an electron with spin $\sigma $ on site $i$, $n_{i\sigma
}\!=\!c_{i\sigma }^{\dagger }c_{i\sigma }$. In the following, we restrict
ourselves to the paramagnetic phase with $n=1$ electron per site at a finite
temperature $T$. One of the cornerstones of strongly correlated electron
systems for this paramagnetic phase is the Mott-Hubbard transition \cite{Mott,Gebhard},
which was analyzed in DMFT in great detail  \cite{DMFT2,DMFTREV,MHT} 
with some -by now settled- controversy for the multi-orbital case  \cite{Liebsch03b,Liebsch03c,Koga04a,Liebsch04a,Koga05a,Koga05b,Inaba05a,Inaba05b,Biermann05b,deMedici05b,Arita06b,Knecht05a,Liebsch05a,vanDongen06a,Ferrero05a,Liebsch06a,Bluemer06a,Liebsch06b}. What is less clear however is how far antiferromagnetic fluctuations above 
the ordering temperature
affect the Mott-Hubbard transition.
For an orientation, see the phase diagram of the Hubbard model in the left inset of Fig.\ \ref{Fig:3D}.
In DMFT, there is no feedback whatsoever of the nearby antiferromagnetic phase onto the
spectrum and the self energy in the paramagnetic phase. If we, for example, completely
suppress the antiferromagnetic phase  through ``frustrating'' the lattice 
with 
longer-range hopping, the DMFT spectrum
is the very same\footnote{except for changes possibly induced by a change of the density of states due the additional hopping}. D$\Gamma$A heals this shortcoming of DMFT.

\begin{figure}[t,b]

\noindent \hspace{-.2cm}\includegraphics[clip=true,width=10.95cm]{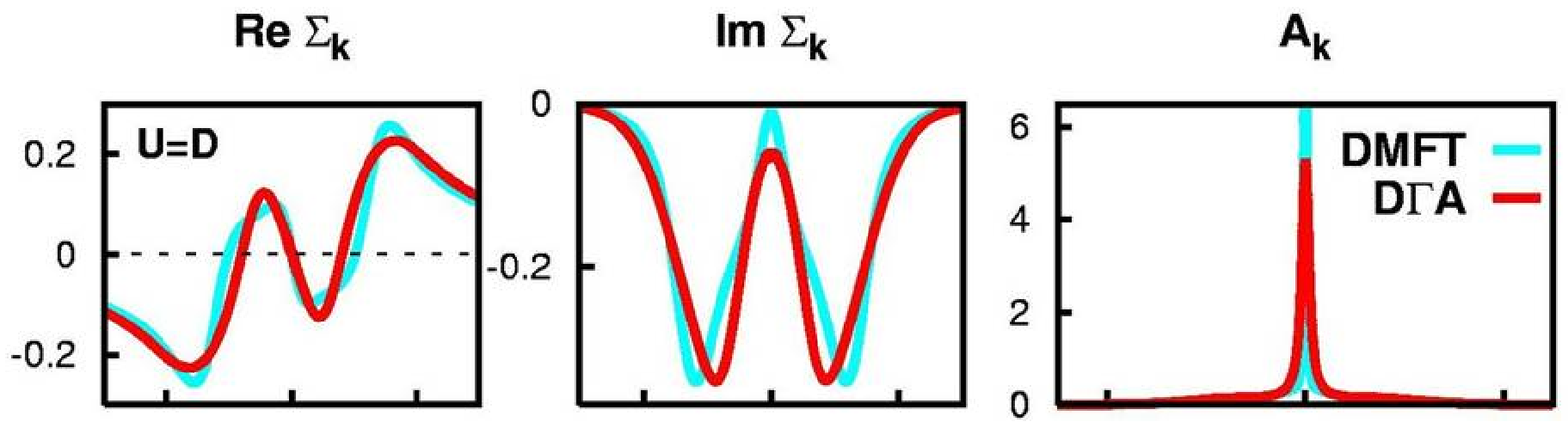} 
\vspace{-1.5cm}

\hspace{10.5cm}{$\leftarrow$}
  \fbox{{\parbox{2.99cm}{
weak damping \\
of QP peak
}}}

\vspace{.7cm}

\noindent\includegraphics[clip=false,width=10.8cm]{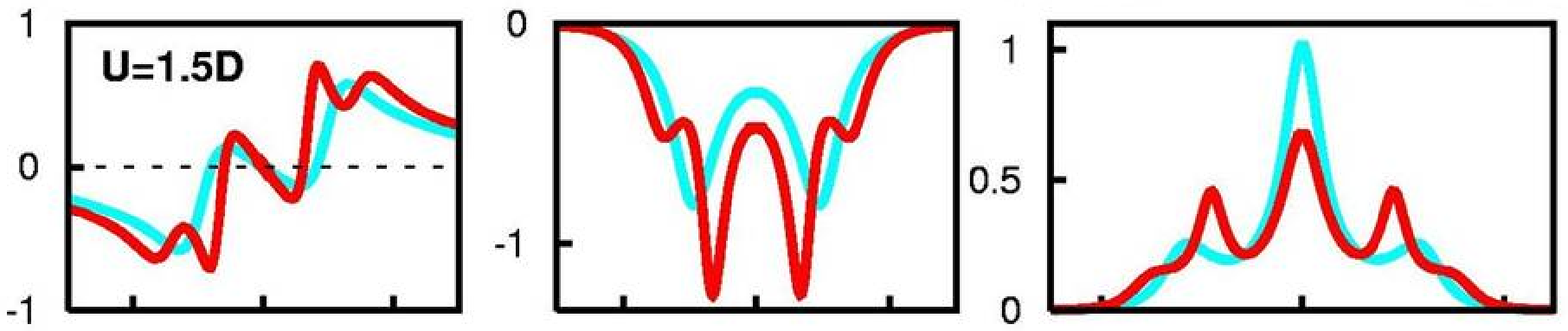} 
\vspace{-1.5cm} 

\hspace{10.5cm} {$\leftarrow$}
  \fbox{{\parbox{2.99cm}{
QP damping\\
strongly enhanced
}}}
\vspace{.7cm} 

\noindent \includegraphics[clip=true,width=10.7cm]{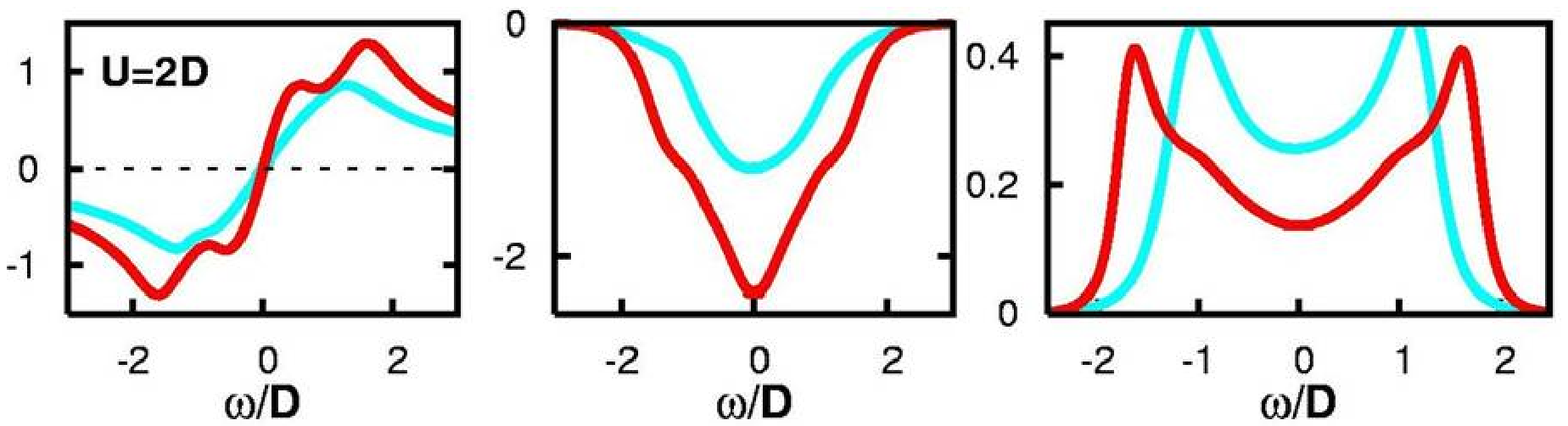} 
\vspace{-2.5cm}

\hspace{10.5cm}{$\leftarrow$}
  \fbox{{\parbox{2.99cm}{
more insulating 
}}}
\vspace{1.7cm}
\caption{D$\Gamma$A self energy (left: real part, middle: imaginary part) and spectral function  (right)  at a point ${\bf k}=(\pi/2,\pi/2,\pi/2)$ on the Fermi surface for
$U\!=\!1D$ (top),  $1.5D$ (central), and $2D$ (bottom) [blue/light grey line]. In the boxes on the right hand side we note the most important
differences in comparison with DMFT [red/dark grey line].  [reproduced from Ref.\ \citen{DGA}]
}
\label{Fig:3D2}
\end{figure}

In our calculations \cite{DGA}, we considered three points in the paramagnetic phase 
of the phase diagram, marked as symbols in left inset of Fig.\ \ref{Fig:3D}.
These points are on the metallic side of the Mott-Hubbard transition and in
the crossover regime where the metal continuously transforms into an insulator.
In DMFT, the metallic side corresponds to a well defined quasiparticle peak (first point considered at $U=1D$, $\beta\equiv 1/T=0.067D$, where $D\equiv2\sqrt{6}t$ is the effective bandwidth/standard deviation of the density of states).
When increasing $U$ to the crossover regime, this quasiparticle peak
 is first strongly damped, i.e, smeared out
 (second point considered at $U=1.5 D$, $\beta=0.089D$; for the DMFT spectrum see Fig.\ \ref{Fig:3D2}).
Note, this change from a metal to a strongly damped quasi-metal 
was confirmed recentely by photoemssion experiments  for  V$_2$O$_3$ by Baldassarre {\em et al.}
\cite{Baldassarre08a}. 
Upon a further increase of $U$ the spectrum develops a pseudogap (third point considered at 
$U=2D$, $\beta=0.1D$). Or, if we take vice versa the insulating gap as a starting point, this insulating gap is filled,
see Ref.\ \citen{Mo04a} for an experimental validation by hands of V$_2$O$_3$.
Increasing $U$ further, eventually  a truly insulating gap is obtained.

In its main panel, Fig.\ \ref{Fig:3d} shows the four-point vertex  irreducible in the spin channel, which
was calculated numerically using exact diagonalization as an impurity solver for the Anderson
impurity model
\cite{DGA}.
As we see, the vertex strongly depends on all three frequencies (main panel and right inset).
It becomes unexpectedly strong if an increasing $U$ enhances  the electronic correlations.
Strong local correlations result in a large vetrex which, in the vicinity of an antiferromagnetic
phase transition, entail strong non-local correlations.

As discussed in Section \ref{Sec:phDGA}, we can calculate the D$\Gamma$A self energy 
and $k$-resolved spectral function from the vertex of 
Fig.\ \ref{Fig:3d}. The results for a wave vector ${\mathbf k}=(\pi/2,\pi/2,\pi/2)$ on
the Fermi surface are presented in Fig. \ref{Fig:3D2}.
On the metallic side at $U=1D$, the non-local antiferromagnetic fluctuations lead
to a relatively weak damping of the quasiparticle peak (QP). Here, the length scales, from
which the main contributions of the antiferromagnetic fluctuations stem, are relatively
long range. Physically, we can understand the damping simply from the fact that the quasiparticles 
live in a bath of bosonic spin fluctuations. The quasiparticle can scatter at these fluctuations,
resulting in a reduced life time, i.e., a larger imaginary part of the self energy.

Entering the crossover regime from the metallic side, the local correlations are strongly enhanced
and entail strong non-local correlations in the vicinity of the antiferromagnetic phase transition.
This can be seen for $U=1.5D$ in Fig. \ref{Fig:3D2}, where we find a strongly damped quasiparticle peak
and a splitting of the Hubbard bands, which can be considered to be a precursor
of the spin polaron side peaks which develop in the antiferromagnetic
phase \cite{Sangiovanni05}.

Finally, in the more insulating-like part of the crossover regime,
at $U=2D$ the antiferromagnetic correlations become even stronger, albeit they are now more 
short range in nature. The paramagnetic phase becomes hence considerably more insulating, indicating a shift
of the Mott-Hubbard transition towards lower values of $U$.
It also shows that the  Mott-Hubbard transition is strongly affected
by  antiferromagnetic fluctuations.

The results shown here are close to the antiferromagnetic phase transition, where the antiferromagnetic fluctuations
are particularly strong. If we increase temperature, i.e., if we are  deep inside the paramagnetic phase, the
difference between D$\Gamma$A and DMFT becomes very small. Already for $T>2T_N$, i.e., twice the
N\'eel temperature, the differences are minor.

All results shown were obtained without $\lambda$ correction; and  in three dimensions
the $\lambda$ correction has only a minor (quantitative) effect. It reduces the antiferromagnetic transition temperature $T_N$ since  $T_N$ is somewhat overestimated in DMFT. But the same quantitative effects as in 
 Fig.\ \ref{Fig:3D2} occur also with $\lambda$ correction albeit at this lower $T_N$, see Ref.\ \citen{DGA2}.

\subsection{Two-dimensional Hubbard model}
\label{Sec:2d}
Let us turn now to the two-dimensional Hubbard model on a square lattice where antiferromagnetic fluctuations are expected to be much more pronounced since the particle-hole ladder diagrams diverges logarithmically in two dimensions. Our insight into these antiferromagnetic fluctuations has been recently improved
not only by the cluster extensions of DMFT dealing with short-range correlations, but also
by the functional renormalization group method
\cite{fRG1,fRG2,fRG3}. The latter is very complementary to the cluster extensions since fRG
deals with correlations on all length-scales, but as a (renormalized)
perturbative approach it does not include the effects of strong correlations such as the Mott-Hubbard transition
and strong renormalizations of the quasiparticle bands. In D$\Gamma$A,  we cover both
of these aspects, strong correlations and multiple length scales, in a single approach.
\begin{figure}[tb]
\begin{center}
\includegraphics[clip=true,width=14.5cm]{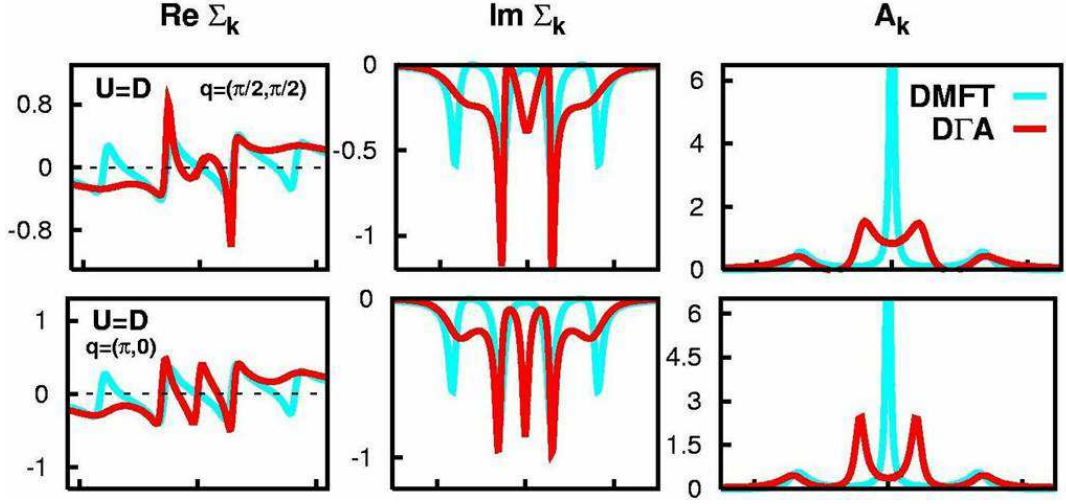} 
\end{center}
\vspace{.2cm}

\caption{Same as in Fig.\ \protect \ref{Fig:3D2} but for the two-dimensional Hubbard model
and a single value of $U=D$ at $\beta=17$. Shown are results for two
different ${\mathbf k}$ points on the Fermi surface, in the nodal (upper panel) and anti-nodal direction (lower panel). The antiferromagnetic fluctuations lead to a complete redistribution of the spectrum in the region of the
quasiparticle peak; a pseudogap opens.
\label{Fig:2D1}}
\end{figure}

Let us emphasize that in two-dimensions the Moriyaesque $\lambda$ correction of Section \ref{Sec:DGAlambda}
 is certainly important to obtain accurate results
since otherwise the antiferromagnetic fluctuations will be overestimated.
Among others, the 
antiferromagnetic transition temperature $T_N$ would be, without $\lambda$ correction,
much too high, i.e., at its DMFT mean field value -- instead of zero as required by the Mermin-Wagner theorem.
Nonetheless, we will present here results without this $\lambda$ correction which can be compared
with the reliable $\lambda$-corrected results of Ref.\ \citen{DGA2}.

Figure \ref{Fig:2D1} shows the same real and imaginary part of the self energy and 
spectral function as  in the first line of Figure \ref{Fig:3D2} for $U=D$, 
but now for two-dimensions (where the effective bandwidth  is $D\equiv 4t$). The difference to three dimensions 
is dramatic since in two-dimensions
the antiferromagnetic fluctuations lead to a complete redistribution of the spectrum.
The quasiparticle peak of DMFT is swallowed up  in a  pseudogap. Even at a relatively weak 
interaction strength $U=D$, the effects of antiferromagnetic fluctuations are much more 
pronounced than for a much stronger interaction strength in three dimensions.
Also the $\mathbf k$-dependence is  dramatic, in contrast to  three dimensions, where 
we  only
showed results for one $\mathbf k$-point because of the weak $\mathbf k$-dependence. Instead, in two dimensions, 
the comparison of nodal [$\mathbf k=(\pi/2,\pi/2)$]
and anti-nodal direction [$\mathbf k=(\pi,0)$] shows a strongly anisotropic pseudogap. 
If we would increase temperatures away from the antiferromagnetic phase transition,
the pseudogap would hence first turn metallic in the nodal direction.

\begin{figure}[tb]
\begin{center}
\includegraphics[clip=true,width=6.9cm]{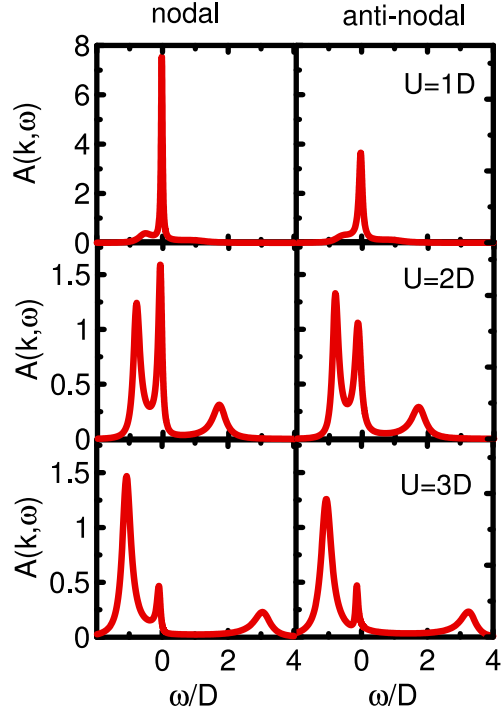} 
\end{center}
\caption{Spectral function for the two-dimensional Hubbard model  off half-filling
in the nodal (left) and anti-nodal (right)
direction. Parameters are  $n\approx 0.8$ electrons/site, a next-nearest neighbor hopping $t'=0.3t$, $\beta=100$ and different values of the interaction
$U=1D$, $2D$, and $3D$ as indicated.
\label{Fig:2D2}}
\end{figure}

Fig.\ \ref{Fig:2D2} presents results off  half-filling for the asymmetric Hubbard model 
with $t'=0.3t$ in a notation where $t$ and $t'$ have opposite prefactors.
At weak coupling, $U=1D$, we clearly see a quasiparticle peak which is already somewhat damped by antiferromagnetic
fluctuations; Hubbard side bands are hardly visible. Note, however the strong anisotropy of the damping which is much stronger in the anti-nodal direction.
Increasing $U$ to $2D$, the damping becomes stronger, but also more isotropic.
Spectral weight is transferred to the Hubbard side bands.
Upon further increasing $U$ to $3D$, these tendencies continue. The quasiparticle weight
is very small; and the corresponding quasiparticle peak is strongly damped in an almost identical way for 
the nodal and anti-nodal direction. As we noted already for the three-dimensional
Hubbard model, the  antiferromagnetic fluctuations for weak coupling are very long-ranged
explaining the pronounced difference between nodal and anti-nodal direction.
The strong non--local correlations at larger $U$ are somewhat more short-range in nature. Consequently, the anisotropy is not so pronounced.

\begin{figure}[tb]
\vspace{.2cm}

\begin{center}
\noindent \includegraphics[clip=true,width=3.5cm]{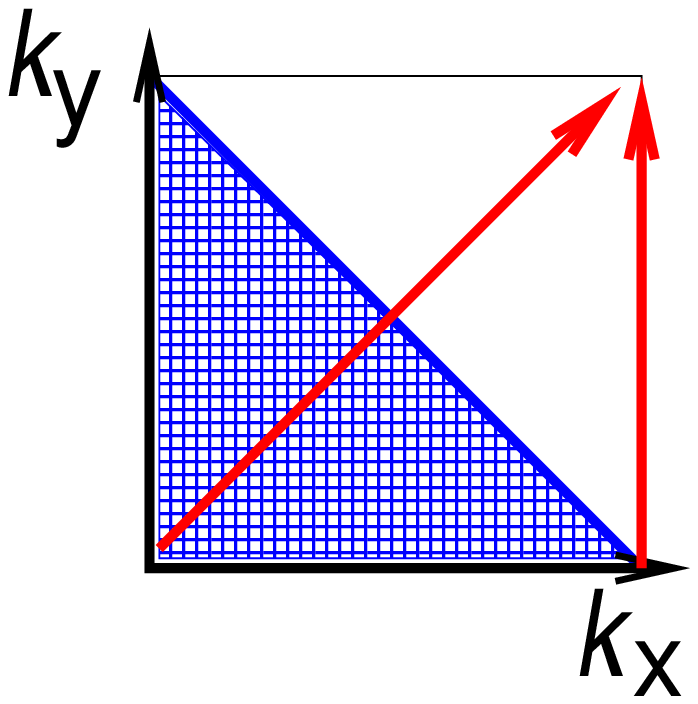}
\end{center}
\vspace{.2cm}

\noindent \includegraphics[clip=true,width=7.cm,height=4.08cm]{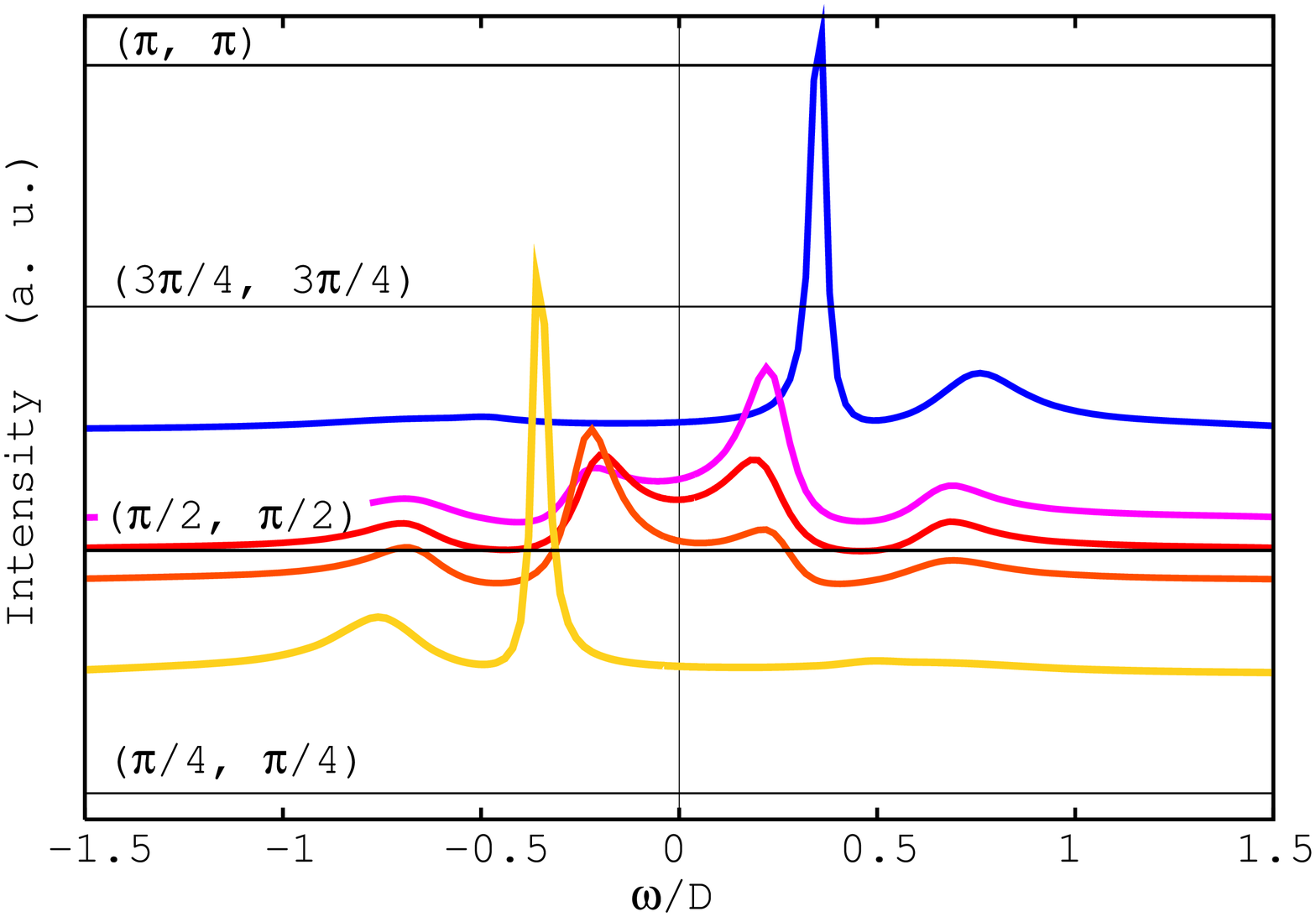}\hspace{.1cm}
 \includegraphics[clip=true,width=7.cm,height=4cm]{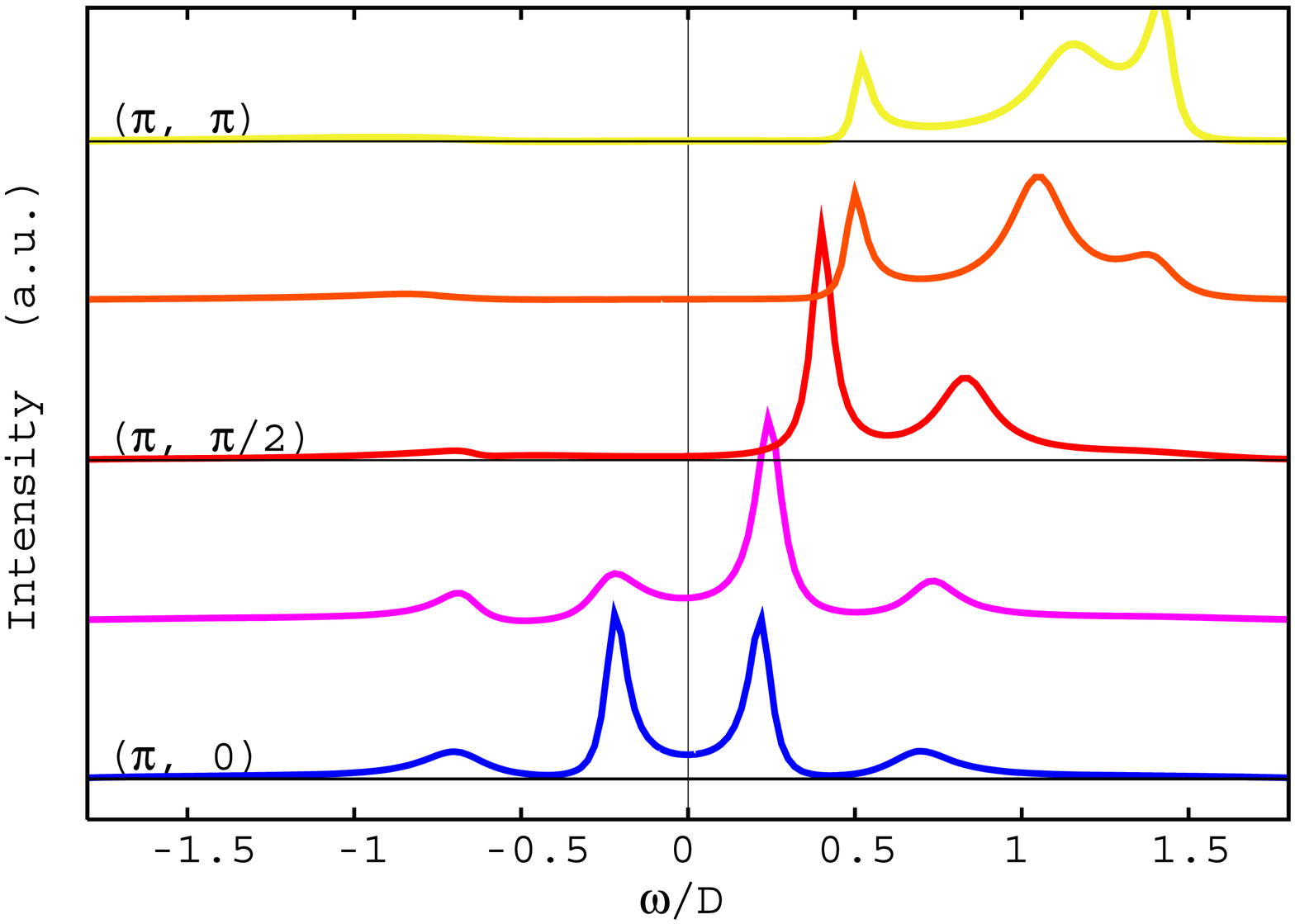}
\vspace{.2cm}

\caption{Spectral function of the two-dimensional Hubbard model 
for different $\mathbf k$-points along the two high symmetry lines indicated in
the upper panel, i.e., in the nodal and anti-nodal direction respectively.
\label{Fig:2D3}}
\end{figure}

For comparison with angular resolved photoemission experiments, we also
provide a scan through the Brillouin zone along lines of high symmetry (in the nodal and anti-nodal direction), see
upper panel of Fig.\ \ref{Fig:2D3}. Note, that $t'=0$ so that the line 
${(0,0) \rightarrow (\pi,0)}$ is connected to the (shown) line  ${(\pi,0) \rightarrow(\pi,\pi)}$
by particle-hole symmetry. When approaching the Fermi surface, the (low-energy) quasiparticle peak
moves towards the Fermi energy. However, this movement stops at some point inducing
a pseudogap. Let us note that with the $\lambda$-correction the antiferromagnetic fluctuations 
are weaker than without $\lambda$ correction \cite{DGA2} but nonetheless still very strong. As a consequence, the pseudogap survives only in the anti-nodal direction.

Most recently,  Li {\em et al.} \cite{Li08}  calculated also the D$\Gamma$A lattice
susceptibility. The results were obtained without $\lambda$-correction
and are similar to those from the dual Fermion approach at higher temperatures
(and much better than DMFT).
However, at lower temperatures there are deviations since without $\lambda$
correction  D$\Gamma$A  overestimates antiferromagnetic fluctuations.

\subsection{One-dimensional Hubbard model}
\label{Sec:1d}
\begin{figure}[tb]
\begin{center}
\includegraphics[clip=true,width=6.4cm,angle=270]{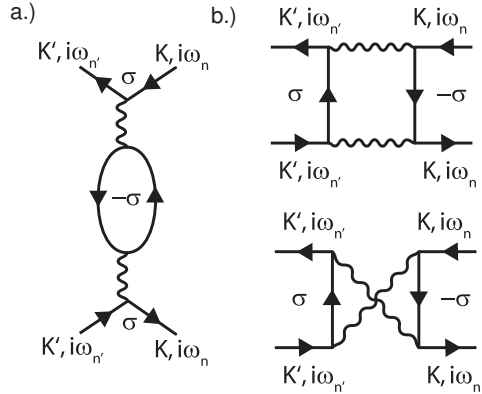} 
\end{center}
\caption{Second order diagrams of the D$\Gamma$Aesque calculation
 by Slezak {\em et al.} for the one-dimensional Hubbard model.  [reproduced from Ref.\ \citen{DGA2}].
\label{Fig:1D1}}
\end{figure}

One of the fascinating effects of electronic correlation is the separation
of  spin and charge in one dimension. Slezak {\em et al.} \cite{DGA2} addressed this
issue in their one-dimensional D$\Gamma$Aesque calculation. As a starting point the authors
considered an 8-site DCA cluster and supplemented it with additional longer range correlation
as in the D$\Gamma$A, albeit with a restriction of the vertex
 to the second order diagrams shown in Fig. \ref{Fig:1D1}.

\begin{figure}[ht]
\begin{center}
\includegraphics[clip=true,width=9.2cm,angle=270]{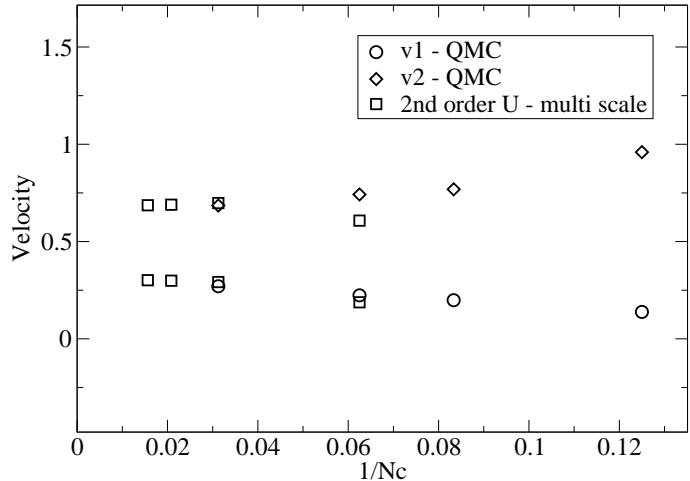} 
\end{center}
\vspace{-1cm}

\caption{Spin and charge velocity as a function of system size $N_c$
calculated by DCA (triangles, circles) and  D$\Gamma$A  with a 8-site cluster as a starting
point (squares). The parameters for the one-dimensional Hubbard model
where $U = W = 1$ ($W$: bandwidth),
$k =\pi/2$, $n = 0.7$ electrons/site, and inverse temperature $\beta= 31$. [reproduced from Ref.\ \citen{DGA2}].
}
\label{Fig:1D2}
\end{figure}

This way, the authors where able to go to a much larger number of lattice sites $N_c$ than in
DCA, see Fig. \ref{Fig:1D2}. This allowed Slezak {\em et al.} to conclude that for $N_c\rightarrow\infty$,
spin and charge velocity are different so that an electron-like spin-charge excitation
separates into its spin and charge components with time. 
This indicates taht the inclusion of
non-local correlations results in a non-Fermi-liquid phase,
suggestively a Luttinger liquid. 

\section{Summary and Outlook}
\label{Sec:summary}
In summary, we have presented the D$\Gamma$A approach in an algorithmic
form, including the restriction to the two particle-hole channels and the
Moriyaesque $\lambda$-correction. Let us also emphasize here that D$\Gamma$A is a complementary
approach to the extended DMFT (EDMFT) \cite{EDMFT1,EDMFT2}. EDMFT deals with the {\em local} correlations 
induced by non-local interactions, whereas D$\Gamma$A deals with the {\em non-local}
correlations stemming from a local interaction.
 The results for the Hubbard model in $d=3$, 2, and 1 dimension
show  effects of antiferromagnetic fluctuations
such as the damping of the quasiparticle peak and the shift of the Mott-Hubbard transition to lower values of $U$ 
in three dimensions,  the opening
of a pseudo-gap and strong anisotropy of the spectrum
in two-dimensions, and the spin-charge separation in one dimension.

Since short- and long-range range antiferromagnetic fluctuations are treated
on the same footing with a  numerical effort  much more manageable than for
cluster extensions of DMFT, we see excellent prospects for this approach in the future.

On a model level, a variety of fascinating phenomena can be addressed such as magnons,
quantum criticality and the interplay of antiferromagnetic fluctuations and superconductivity.
But also realistic multi-orbital calculations are possible for which the cluster extensions
are too severely restricted because of the numerical effort. Such local density approximation (LDA) + D$\Gamma$A
calculations would, among others, 
allow us to study the effect of antiferromagnetic fluctuations on the Mott-Hubbard transition in
V$_2$O$_3$ where the model scenario of Section \ref{Sec:3d}
is realized in an actual material and which was previously studied successfully  in LDA+DMFT \cite{Held01a,Keller04a,Laad03b,Laad06a,Poteryaev07}.
We hope that this review, along the lines of a talk given at the  Yukawa Institute in November 2007,
will stimulate more groups to apply the D$\Gamma$A approach to challenging problems 
of strongly correlated electron systems.

\section*{Acknowledgements}
 We thank the Yukawa Institute in Kyoto, Japan, for its generous hospitality and
W.\ Metzner, M. Capone, C.\ Castellani,
G.\ Sangiovanni, and R.\ Arita for stimulating discussions; we are indebted
to M. Capone also for providing the DMFT(ED) code which has served as a
starting point. This work was supported by
 the Russian Basic Research Foundation through Grant
No.~747.2003.2 (AK).

%


\begin{thebibliography}{99}
  
\bibitem{DMFT}
W.~Metzner and D.~Vollhardt,
\newblock {Phys. Rev. Lett. \bf 62}   (1989),  324.


\bibitem{DMFT2}
A.~Georges and G.~Kotliar, \newblock Phys. Rev. B {\bf 45} (1992), 6479.


\bibitem{DMFTREV}
 A.~Georges, G.~Kotliar, W.~Krauth, and M.~Rozenberg,
Rev. Mod. Phys. {\bf 68}  13 (1996).


\bibitem{DMFTPT}
G.~Kotliar and D.~Vollhardt, \newblock Physics Today {\bf March} 53 (2004).


\bibitem{MHT}
R.~Bulla, \newblock Phys. Rev. Lett. {\bf 83} (1999), 136.

\bibitem{MAG1}
M.~Jarrell, \newblock Phys. Rev. Lett. {\bf 69} (1992), 168.

\bibitem{MAG2}
K.~Held, M.~Ulmke, N.~Bl{\"{u}}mer and D.~Vollhardt, \newblock Phys. Rev. B {\bf 56} (1997), 14469.

\bibitem{MAG3}
M.~Ulmke, \newblock Eur. Phys. J. B {\bf 1},  (1998) 301;

\bibitem{MAG4}
N.~Furukawa, \newblock J. Phys. Soc. Jap. {\bf 63} (1994), 3214.

\bibitem{MAG5}
K.~Held and D.~Vollhardt, \newblock Euro. Phys. J. B {\bf 5} (1998).

\bibitem{kink1}
I.~A. Nekrasov, K.~Held, G.~Keller, D.~E. Kondakov, T.~Pruschke, M.~Kollar,
  O.~K. Andersen, V.~I. Anisimov and D.~Vollhardt, \newblock Phys. Rev. B {\bf 73}, 155112 (2006).

\bibitem{kink2}K.~Byczuk, M.~Kollar, K.~Held, Y.-F. Yang, I.~A. Nekrasov, T.~Pruschke and
  D.~Vollhardt, \newblock Nature Physics {\bf 3}
 168  (2007).

\bibitem{kink3}A. Toschi, M. Capone, C. Castellani, and
K. Held, arXiv:0712.3723. 

\bibitem{LDADMFT1}
V.~I. Anisimov, A.~I. Poteryaev, M.~A. Korotin, A.~O. Anokhin and G.~Kotliar, \newblock J. Phys. Cond. Matter {\bf 9} (1997),  7359.

\bibitem{LDADMFT2}
A.~I. Lichtenstein and M.~I. Katsnelson, 
\newblock Phys. Rev. B {\bf 57} 198, 6884 (1998).



\bibitem{LDADMFT3}
K.~Held, I.~A. Nekrasov, G.~Keller, V.~Eyert, N.~Bl\"umer, A.~McMahan,
  R.~Scalettar, T.~Pruschke, V.~I. Anisimov and D.~Vollhardt,
phys. stat. sol. (B) {\bf 243} 2599 (2006).

\bibitem{LDADMFT4}
G.~Kotliar, S.~Y. Savrasov, K.~Haule, V.~S. Oudovenko, O.~Parcollet and C.~A.
  Marianetti, \newblock Rev. Mod. Phys.  {\bf 78}, 865 (2006).


\bibitem{LDADMFT5}
K. Held, Adv. Phys. {\bf 56}, 829 (2007). 

\bibitem{review}
T.\ A.\ Maier, M.\ Jarrell, T.\ Pruschke, M.\ H.\ Hettler,
Rev.\ Mod.\ Phys.\ {\bf 77}, 1027 (2005).

\bibitem{clusterDMFT}G.\ Kotliar, S. Y. Savrasov1, G. P\'alsson, and G. Biroli, Phys.\ Rev.\ Lett.\ {\bf 87}, 186401 (2001).


\bibitem{clusterLK}A.\ I.\ Lichtenstein
and M.\ I.\ Katsnelson, Phys.\ Rev. B {\bf 62}, R9283 (2000).
\bibitem{clusterDMFT2}
M.~Capone and G.~Kotliar Phys. Rev. B {\bf 74}, 054513 (2006).


\bibitem{DCA1} M.\ H.\ Hettler,
A.\ N.\ Tahvildar-Zadeh, M.\ Jarrell,
T.\ Pruschke, and H.\ R.\ Krishnamurthy,
Phys. Rev. B {\bf 58}, (1998) 7475.
\bibitem{DCA2}
C. Huscroft, M. Jarrell, Th. Maier, S. Moukouri, and A. N. Tahvildarzadeh, Phys. Rev. Lett. {\bf 86}, 139 (2001).

\bibitem{DCA3}
Cluster sizes can be extended by using
 the fluctuation exchange approximation \cite{FLEX}
see
K. Aryanpour, M. H. Hettler, and M. Jarrell,
Phys. Rev. B {\bf 67}, 085101 (2003).

\bibitem{FLEX}
  N.\ E.\ Bickers,    D.\ J.\ Scalapino, and S.\ R.\ White,
 Phys. Rev. Lett. {\bf 62}, 961 (1989).

\bibitem{Potthoff03a}
M.~Potthoff, \newblock Eur. Phys. J. B {\bf 32} 429 (2003).

\bibitem{Potthoff06a}
M.~Potthoff and M.~Balzer Phys. Rev. B {\bf 75}, 125112 (2007).

\bibitem{Schiller}
A.~Schiller and K.~Ingersent, \newblock Phys. Rev. Lett. {\bf 75} 113 (1995).

\bibitem{Zarand00}
G.~Zar\'and, D.~L. Cox and A.~Schiller, \newblock Phys. Rev. B {\bf 62} 16227 (R) (2000).

\bibitem{Capone04a}
M.~Capone, M.~Civelli, S.~S. Kancharla, C.~Castellani and G.~Kotliar, \newblock Phys. Rev. B {\bf 69} 195105 (2004).

\bibitem{clusterDMFT3}
E. Koch, G. Sangiovanni, and O. Gunnarsson, arXiv:0710.1247.


\bibitem{DCASC}
T.~Maier, M.~Jarrell, T.~Pruschke and J.~Keller, \newblock Phys. Rev. Lett. {\bf 85} 1524 (2000).

\bibitem{Maier05a}
T.~A. Maier, M.~Jarrell, T.~C. Schulthess, P.~R.~C. Kent and J.~B. White, \newblock Phys. Rev. Lett. {\bf 95} 237001 (2005).

\bibitem{Arita06a}
R.~Arita and K.~Held, \newblock Phys. Rev. B {\bf 73} 064515 (2006).

\bibitem{Poteryaev04a}
A.~I. Poteryaev, A.~I. Lichtenstein and G.~Kotliar, \newblock Phys. Rev. Lett. {\bf 93} 086401 (2004).

\bibitem{Mazurenko02a}
V.~V. Mazurenko, A.~I. Lichtenstein, M.~I. Katsnelson, I.~Dasgupta,
  T.~Saha-Dasgupta and V.~I. Anisimov, \newblock Phys. Rev. B {\bf 66} 081104 (R) (2002).

\bibitem{Biermann05a}
S.~Biermann, A.~Poteryaev, A.~I. Lichtenstein and A.~Georges, \newblock Phys. Rev. Lett. {\bf 94} 026404 (2005).

\bibitem{BSS}
R.~Blankenbecler, D.~J. Scalapino and R.~L. Sugar, \newblock Phys. Rev. D {\bf 24} 2278 (1981).

\bibitem{Sadovskii05}E.\ Z.\ Kuchinskii {\sl et al.}, Sov.\ Phys.\ JETP Lett. {\bf 82}, 98 (2005).

\bibitem{Sadovskii05b}
M.\ V.\ Sadovskii {\em et al.}, Phys. Rev. B {\bf 72}, 155105 (2005).

\bibitem{DualFermion}
 A. N. Rubtsov, M. I. Katsnelson, and A. I. Lichtenstein, Phys. Rev. B {\bf 77}, 033101 (2008)); S. Brener, H. Hafemann, A. N. Rubtsov, M. I. Katsnelson, and A. I. Lichtenstein, Phys. Rev. B {\bf 77}, 195105 (2008).

\bibitem{Tokar07a} V. I. Tokar and R. Monnier,  cond-mat/0702011 (unpublished).

\bibitem{DGA} A. Toschi, A. A. Katanin, and K. Held,
 Phys. Rev. B 75, 045118 (2007).

\bibitem{DGA3} A. A. Katanin, A. Toschi, and K. Held, (unpublished).

\bibitem{DGA2} C. Slezak, M. Jarrell, Th. Maier,  and J. Deisz, cond-mat/0603421 (unpublished).

\bibitem{Kusunose}
 H.\ Kusunose,   J.\ Phys.\ Soc.\ JPN.\ {\bf 75}, 054713 (2006).

\bibitem{Hirsch86a}
J.~E. Hirsch and R.~M. Fye, \newblock Phys. Rev. Lett. {\bf 56} 2521 (1986).

\bibitem{Feldbacher06}
M.~Feldbacher, K.~Held and F.~F. Assaad, \newblock Phys. Rev. Lett. {\bf 96} 139702 (2006).
\bibitem{Rubtsov04a}
A.~N. Rubtsov and A.~I. Lichtenstein, \newblock JETP Lett. {\bf 80} 61 (2004).

\bibitem{Werner06a}
P.~Werner and A.~J. Millis, \newblock Phys. Rev. B {\bf 74} 155107 (2006).

\bibitem{Sakai06a}
S.~Sakai, R.~Arita, K.~Held and K.~Aoki, \newblock Phys. Rev. B {\bf 74} 155102 (2006).

\bibitem{Dzy} Y. A. Bychkov, L. P. Gor'kov, and I. E. Dzyaloshinskii, Zh. Exp. Teor. Fiz. {\bf 50}, 738 (1966)
[Sov. Phys. JETP {\bf 23}, 489 (1966)].

\bibitem{parquet} N.\ E.\ Bickers and S.\ R.\ White, Phys.\ Rev.\ B {\bf 43}, 8044 (1991).

\bibitem{Janis2} V.\ Jani\v{s}, Phys. Rev. B {\bf 60} 11345 (1999).



\bibitem{Moriya}
T.\ Moriya,  ``Spin fluctuations in Itinerant Electron Magnetism''  (Springer, 1985).



\bibitem{Mott}
 N. F. Mott, Rev.\ Mod.\ Phys.\ {\bf 40}, 677
(1968); {\sl Metal-Insulator Transitions} (Taylor \& Francis,
London, 1990)

\bibitem{Gebhard}
 F. Gebhard, {\sl The Mott Metal-Insulator
Transition} (Springer, Berlin, 1997).

\bibitem{Liebsch03b}
A.~Liebsch, \newblock Europhys. Lett. {\bf 63} 97 (2003).

\bibitem{Liebsch03c}
A.~Liebsch, \newblock Phys. Rev. Lett. {\bf 91} 226401 (2003).

\bibitem{Koga04a}
A.~Koga, N.~Kawakami, T.~M. Rice and M.~Sigrist, \newblock Phys. Rev. Lett. {\bf 92} 216402 (2004).

\bibitem{Liebsch04a}
A.~Liebsch, \newblock Phys. Rev. B {\bf 70} 165103 (2005).

\bibitem{Koga05a}
A.~Koga, K.~Inaba and N.~Kawakami, \newblock Prog. Theo. Phys. Suppl. {\bf 160} 253 (2005).

\bibitem{Koga05b}
A.~Koga, N.~Kawakami, T.~M. Rice and M.~Sigrist, \newblock Phys. Rev. B {\bf 72} 045128 (2005).

\bibitem{Inaba05a}
K.~Inaba, A.~Koga, S.~I. Suga and N.~Kawakami, \newblock J. Phys. Soc. Jap. {\bf 74} 2393 (2005).

\bibitem{Inaba05b}
K.~Inaba, A.~Koga, S.~I. Suga and N.~Kawakami, \newblock Phys. Rev. B {\bf 72} 085112 (2005).

\bibitem{Biermann05b}
S.~Biermann, L.~de' Medici and A.~Georges, \newblock Phys. Rev. Lett. {\bf 95} 206401 (2005).

\bibitem{deMedici05b}
L.~de' Medici, A.~Georges and S.~Biermann, \newblock Phys. Rev. B {\bf 72} 205124 (2005).

\bibitem{Arita06b}
R.~Arita and K.~Held, \newblock Phys. Rev. B {\bf 72} 201102(R) (2005).

\bibitem{Knecht05a}
C.~Knecht, N.~Bl{\"u}mer and P.~G.~J. van Dongen, \newblock Phys. Rev. B {\bf 72} 081103 (R) (2005).

\bibitem{Liebsch05a}
A.~Liebsch, \newblock Phys. Rev. Lett. {\bf 95} 116402 (2005).

\bibitem{vanDongen06a}
P.~G.~J. van Dongen, C.~Knecht and N.~Bl{\"u}mer, \newblock Phys. Stat. Sol. B {\bf 243} 116 (2006).

\bibitem{Ferrero05a}
M.~Ferrero, F.~Becca, M.~Fabrizio and M.~Capone, \newblock Phys. Rev. B {\bf 72} 205126 (2005).

\bibitem{Liebsch06a}
A.~Liebsch and T.~A. Costi, Euro. Phys. J. B {\bf 51}, 523 (2006).

\bibitem{Bluemer06a}
N.~Bl\"umer, C.~Knecht, K.~Pozgajcic and P.~van Dongen  (2006), \newblock {cond-mat/0609758}.

\bibitem{Liebsch06b}
A.~Liebsch  (2006), \newblock {cond-mat/0610482}.


\bibitem{Mo04a}S.-K. Mo, H.-D. Kim, J. W. Allen, G.-H. Gweon, J. D. Denlinger, J.-H. Park, A. Sekiyama, A. Yamasaki, S. Suga, P. Metcalf, and K. Held, \newblock Phys. Rev. Lett. {\bf 93} 076404 (2004).

\bibitem{Baldassarre08a}
L. Baldassarre, A. Perucchi, D. Nicoletti, A. Toschi, G. Sangiovanni, K. Held, M. Capone, M. Ortolani, L. Malavasi, M. Marsi, P. Metcalf, P. Postorino, and S. Lupi, Phys. Rev. B {\bf 77}, 113107 (2008)

\bibitem{Sangiovanni05}
G.~Sangiovanni, A.~Toschi, E.~Koch, K.~Held, M.~Capone, C.~Castellani,
  O.~Gunnarsson, S.-K. Mo, J.~W. Allen, H.-D. Kim, A.~Sekiyama, A.~Yamasaki,
  S.~Suga and P.~Metcalf, \newblock Phys. Rev. B {\bf 73} 205121 (2006).


\bibitem{fRG1} D.\ Zanchi and H.\ J.\ Schulz, Phys. Rev. B {\bf 54},
9509 (1996).

\bibitem{fRG2}
  C.\ J.\ Halboth and W.\ Metzner, Phys.\ Rev.\ Lett. {\bf 85},
5162 (2000); C.\ Honerkamp {\sl et al.}, Phys.\ Rev.\ B {\bf 63}, 035109 (2001); Phys.\ Rev.\ Lett.\ {\bf 87}, 187004 (2001).

\bibitem{fRG3}
A.\ A.\ Katanin and A.\ P.\ Kampf, Phys.\ Rev.\ B {\bf 68}, 195101 (2003).


\bibitem{Li08}
 G. Li, H. Lee, and H. Monien, arXiv:0804.3043.
\bibitem{EDMFT1} Q. Si and J.L. Smith, Phys. Rev. Lett. {\bf 77}, 3391 (1996).

\bibitem{EDMFT2}
 H. Kajueter, Ph.D. thesis, Rutgers University (1996).

\bibitem{Held01a}
K.~Held, G.~Keller, V.~Eyert, V.~I. Anisimov and D.~Vollhardt, \newblock Phys. Rev. Lett. {\bf 86} 5345 (2001).

\bibitem{Keller04a}
G.~Keller, K.~Held, V.~Eyert, D.~Vollhardt and V.~I. Anisimov, \newblock Phys. Rev. B {\bf 70} 205116 (2004).


\bibitem{Laad03b}
M.~S. Laad, L.~Craco and E.~M\"uller-Hartmann, \newblock Phys. Rev. Lett. {\bf 91} 156402 (2003).

\bibitem{Laad06a}
M.~S. Laad, L.~Craco and E.~M{\"u}ller-Hartmann, \newblock Phys. Rev. B {\bf 73} 045109 (2006).

\bibitem{Poteryaev07}
A.I. Poteryaev, J.M. Tomczak, S. Biermann, A. Georges, A.I. Lichtenstein, A.N. Rubtsov, T. Saha-Dasgupta and O.K. Andersen, Phys. Rev. B {\bf 76}, 085127 (2007).

\end{thebibliography}
\end{document}